\documentclass[showpacs]{revtex4}
\usepackage[english]{babel}
\usepackage{amsmath}

\usepackage{latexsym}

\usepackage{epsfig}

\begin{document}

\title{Probabilistic approach to a proliferation and migration dichotomy
in the tumor cell invasion}

\author{Sergei Fedotov${}^1$ and Alexander Iomin${}^2$ }

\affiliation{${}^1$ Department of Mathematics, University of Manchester,
Manchester M60 1QD, UK \\
${}^2$ Department of Physics, Technion, Haifa, 32000, Israel}

\date{\today}
\begin{abstract}

The proliferation and migration dichotomy of the tumor cell invasion
is examined within a two-component continuous time random walk
(CTRW) model. The balance equations for the cancer cells of two
phenotypes with random switching between cell proliferation and
migration are derived. The transport of tumor cells is formulated
in terms of the CTRW with an arbitrary waiting time distribution
law, while proliferation is modelled by a logistic growth. The
overall rate of tumor cell invasion for normal diffusion and
subdiffusion is determined.

\end{abstract}


\maketitle

\section{Introduction}

One of the main features of malignant brain cancer is the ability
of tumor cells to invade the normal tissue away from the
multi-cell tumor core. Invasion of healthy tissue by a solid tumor
(the core), and the role of oxygen and nutrient delivery have been
the subject of extensive studies reflected in modern surveys, see
e.g., \cite{BelMaini}. Experimental data for a glioma cancer show
that the proliferation rate of migratory cells is lower in the
invasion region than in the core.  It turns out that the
proliferation and migration of cells are mutually exclusive: the
high motility suppresses cell proliferation and vice versa. This
phenomenon is known as the \textit{migration-proliferation
dichotomy} \cite{Giese1,Giese2}. The exact mechanism of switching
between the two phenotypes of glioma cells is not known. There are
several phenomenological models for this dichotomy. One can assume
that the diffusion coefficient of cancer cells is a decreasing
function of cell density \cite{khain1}. As a result the cancer
cell motility is greater in the invasion zone because of the small
density of cells there. One can also assume the dependence of the
proliferation term on cell density such that the proliferation
rate increases with density \cite{khain2}. An interesting
dynamical model for the phenotype switch was suggested in
\cite{athale}. However, this mathematical model involves many
parameters, some of which are difficult to estimate. Recently the
authors proposed a stochastic approach for the
proliferation-migration switching that involves only two
parameters \cite{fi07}. The transport process was formulated in
the framework of the continuous time random walk (CTRW)
\cite{shlezinger,klafter,iom1}. The main reason for employing the
CTRW model was to give the mesoscopic description of cancer cell
motility in terms of the random jump distribution and waiting
times. One of the main purposes was to take into account anomalous
transport (subdiffusion) leading to slow motility of cancer cells
in the invasive zone. Among all possible cancer cell genotypes,
leading to six main alternations of malignant growth
\cite{HananWeinberg}, cell motility and invasion are most
important for our consideration. The standard diffusion
approximation for the transport (which is the parabolic limit of
kinetics) together with a logistic growth yields an overestimation
of the overall growth \cite{F1,FM}. Since the motility is the most
critical feature of brain cancer, causing treatment failure, there
is a need for a proper description of cancer cell motility beyond
the standard diffusion approximation. In this connection, the
hyperbolic limit of the multi-cellular microscopic system is
important \cite{bbns2007} to take into account cellular
interaction in the description of macroscopic dynamics. A very
interesting agent-based model was developed \ recently by Mansury
and Deisboeck \cite{mansury1}. The transport process is described
in terms of the local-search mechanism performed by tumor cells.
The purpose of this ``conscious'' search is to find and then
invade the most permissive location in extracellular matrix. A
simplified scheme of migration--proliferation dichotomy in terms
of CTRW was considered in \cite{iom1,iomin}. It involves two
steps: cell fission with the characteristic time $\mathcal{T}_{f}$
and cell transport with duration $\mathcal{T}_{t}$. During the
time scale $\mathcal{T}_{f}$, the cells interact strongly and
motility of the cells is small. During the time $\mathcal{T}_{t}$,
interaction between the cells is weak and motility of the cells is
determined by a ``jump'' length $\sim\mathcal{T}_{t}$.

Cell invasion is a very complex process controlled by matrix
adhesion (see review \cite{Giese1}). It involves several steps
including receptor-mediated adhesion of cells to extracellular
matrix (ECM), matrix degradation by tumor-secreted proteases
(proteolysis), detachment from ECM adhesion sites, and active
invasion into intercellular space created by protease degradation.
One of the purposes of this paper is to give a description of this
complicated cell transport in terms of a non-symmetrical random
walk model with memory effects. Chemotaxis and haptotaxis are
taken into account by the biased random walk of cells that respond
to external signals without alteration and migrate away from the
tumor core. Matrix adhesion effects are modelled by using the
heavy-tailed waiting time distributions that lead to subdiffusion
of tumor cells.

\section{Two-component CTRW with proliferation}

\subsection{Balance equations}

In this paper we present a detailed analysis of the migration and
proliferation of glioma cells in the framework of a two-component
continuous time random walk with proliferation. The paper is an
essential extension of our Letter \cite{fi07} with new results and
examples. Based on experimental observations of
\textit{migration-proliferation dichotomy}, we assume that the
process of tumor cell invasion consists of two states. In state
$1$ (migratory phenotype) the cells randomly move but there is no
cell proliferation. In state $2$ (proliferating phenotype) the
cancer cells do not migrate and only proliferation takes place. To
describe the random switching between the two phenotypes, we
employ the two-state Markov chain model. The cell of type $1$
remains in state $1$ during a waiting time $\tau _{1}$ and then
switches to a cell of type $2$. After a waiting time $\tau _{2}$,
spent in state $2$, it switches back to a cell of type $1$. Both
waiting times $\tau _{1}$ and $\tau _{2}$ are mutually independent
random variables exponentially distributed with parameters $\beta
_{1}$ and $\beta _{2}$:
\begin{equation}\label{swit}
{\cal P}(\tau _{k})=\beta _{k}\exp \left( -\beta _{k}\tau
_{k}\right) \;\;\;k=1,2.
\end{equation}
Here the parameters $\beta _{k}$ are the switching rates, namely,
$\beta_{1} $ is the switching rate from state $1$ to $2$, while
$\beta _{2}$ determines the transition rate $2\rightarrow 1$. Note
that the generalization for the renovation processes with
arbitrary probability densities for switching times is
straightforward. An important feature of the present analysis is
an observation of the influence of the migration-proliferation
dichotomy on the overall invasion rate of cancer cells. In what
follows we show how the overall propagation rate $u$ depends on
the parameters $\beta _{k}$.

We consider the growing tumor spheroid consisting of the tumor
core with a high density of cells and the outer invasive zone
where the cell density is much smaller. To describe the cancer
cells of the two phenotypes we introduce the density of the cells
of migrating phenotype, $n_{1}(t,\mathbf{x})$, and the density of
the cells of proliferating phenotype, $n_{2}(t,\mathbf{x})$. The
balance equations for $n_{1}(t,\mathbf{x})$ and
$n_{2}(t,\mathbf{x})$ are
\begin{eqnarray}
n_{1}(t,\mathbf{x}) &=&n_{1}(0,\mathbf{x})\Psi (t)e^{-\beta
_{1}t}+\int_{0}^{t}\int_{R^{d}}n_{1}(t-s,\mathbf{x}-\mathbf{z})\Phi (s,%
\mathbf{z})e^{-\beta _{1}s}d\mathbf{z}ds  \notag \\
&&+\beta _{2}\int_{0}^{t}n_{2}(t-s,\mathbf{x})\Psi (s)e^{-\beta
_{1}s}ds\,, \label{n1}
\end{eqnarray}
\begin{eqnarray}
n_{2}(t,\mathbf{x}) &=&n_{2}(0,\mathbf{x})e^{-\beta
_{2}t}+\int_{0}^{t}f\left( n_{1}(t-s,\mathbf{x}),n_{2}(t-s,\mathbf{x}%
)\right) e^{-\beta _{2}s}ds  \notag \\
&&+\beta _{1}\int_{0}^{t}n_{1}(t-s,\mathbf{x})e^{-\beta_{2}s}ds\,,
\label{n2}
\end{eqnarray}
where $\Phi (s,\mathbf{z})$ is the joint probability density
function of making a jump\textbf{\ }$\mathbf{z}$ in the time
interval $s$ to $s+ds$, and $R_{d}$ denotes the integration is
over $d$-dimensional space. The one dimensional case ($d=1$) was
considered in \cite{fi07}.

Cell migration (random jumps) involves a receptor-mediated
adhesion to matrix proteins, matrix degradation by proteases,
detachment from adhesion sites, active invasion into ``new''
intercellular space formed by degradation, etc. It would be
extremely difficult to build up a rigorous deterministic model for
this process. Since these factors are too many, we believe that a
good alternative to such a model is a random walk with memory
effects. The active mechanism of migration of tumor cells involves
small random jumps and delay time between jumps. The latter might
be of the same order as the proliferation time. This dynamics is
obviously random and its distribution is given by the probability
density function (pdf) $\psi (s):$
\begin{equation}
\psi (s)=\int_{R^{d}}\Phi (s,\mathbf{z})d\mathbf{z}\, ,
\end{equation}
where $\Phi (s,\mathbf{z})$ is the joint pdf.

 Equation (\ref{n1}) is the conservation law for cells of type $1$
at time $t$ at position $\mathbf{x}.$ The first term on the right
hand side $n_{1}(0,\mathbf{x})\Psi (t)e^{-\beta _{1}t}$ represents
cells of type $1$\ that stay up to time $t$ at position
$\mathbf{x}$ such that no jump occurred, and no switch took place.
This term involves the function $\Psi (t)$
\begin{equation}
\Psi (t)=1-\int_{0}^{t}\psi (s)ds  \label{Psi}
\end{equation}
which is the probability that a cell of type $1$ makes no jump
until time $t$ . Note that the exponential factor
\begin{equation*}
e^{-\beta _{k}t}=1-\int_{0}^{t}{\cal P}(\tau _{k})ds\,,\;\;\;k=1,2
\end{equation*}
is the probability that cells of phenotypes $k$ do not switch
until time $t$. The independence of the random jumps and switching
gives us the probability $\Psi (t)e^{-\beta _{1}t}$ while the
first factor $n_{1}(0, \mathbf{x})$ is the initial density of
cells of type $1$ at $\mathbf{x}$.

The second term
\begin{equation*}
\int_{0}^{t}\int_{R^{d}}n_{1}(t-s,\mathbf{x}-\mathbf{z})\Phi
(s,\mathbf{z} )e^{-\beta _{1}s}d\mathbf{z}ds
\end{equation*}
gives us the number of cells of type $1$ arriving at $\mathbf{x}$
up to time $t$. We assume the following random mechanism of
migration: the cell of type $1$ at time $t-s$ at position
$\mathbf{x}-\mathbf{z}$ waits a random time $s$ before jumping a
distance $\mathbf{z}$ at position $\mathbf{x}$ and remains a cell
of type $1$. The last term
\begin{equation*}
\beta _{2}\int_{0}^{t}n_{2}(t-s,\mathbf{x})\Psi (s)e^
{-\beta_{1}s}ds
\end{equation*}
represents the number of cells of type $2$ that switch to the cell
of type $1$ up to time $t$ and remain the cells of type $1$ (the
factor $e^{-\beta _{1}s}$). It also takes into account the fact
that if transition $2\rightarrow 1$ happens at time $t-s$, then no
jump takes place during the remaining time $s$ (the factor $\Psi
(s)$).

Equation (\ref{n2}) describes the balance of cells of
proliferating phenotype (no jumps). The first term on the right
hand side, $n_{2}(0,\mathbf{x})e^{-\beta _{2}t}$, is the density
of cells of type $2$\ that stay up to time $t$ at position
$\mathbf{x}$ such that no switch $2\rightarrow 1$ takes place. The
second term on the right hand side
\begin{equation*}
\int_{0}^{t}f\left(
n_{1}(t-s,\mathbf{x}),n_{2}(t-s,\mathbf{x})\right) e^
{-\beta_{2}s}ds
\end{equation*}
is the proliferation rate for cell of type $2$, which occurs
providing that no switch takes place up to time $t$. The last term
\begin{equation}
\beta _{1}\int_{0}^{t}n_{1}(t-s,\mathbf{x})e^{-\beta _{2}s}ds
\end{equation}
gives the number of cell of type $1$ switching to the state $2$
over the time interval $(0,t)$.

It is well known that the CTRW modelling is a standard technique
for studying anomalous diffusion \cite{shlezinger,klafter}. We
employ this technique to take into account subdiffusion that leads
to slow motility of cancer cells in the invasive zone. In this
paper each random step of a cancer cell is characterized by a
waiting time $s$ and a jump\textbf{\ }$\mathbf{z}$ which are
distributed according to the joint pdf $ \Phi (s,\mathbf{z})$.
This pdf can be written in a decoupled form
\begin{equation}
\Phi (s,\mathbf{z)}=\psi (s)\rho (\mathbf{z}),  \label{dec}
\end{equation}
where $\psi (s)$ is waiting time pdf and $\rho (\mathbf{z})$ is
the pdf of cell jumps. This form corresponds to the case when the
random waiting time and the individual displacement are
independent. The subdiffusion regime occurs when the mean waiting
time $<t >=\int_{0}^{\infty }\tau \psi (\tau )d\tau $ is infinite
and the spherically symmetrical pdf $\rho(|\mathbf{x|})=\rho(r)$
has a finite variance $\sigma ^{2}=\int r^{2}\rho(r)dr<\infty $,
where $r$ is the radius of the spheroid. If the asymptotic
behavior for the waiting-time density $\psi $ $(t )$ for\ large $t
$ is $t ^{-1-\zeta }$ with $ 0<\zeta <1$, the mean waiting time
$<t>$ is infinite and the mean-square displacement $\sigma
^{2}t^{\zeta }$ corresponds to subdiffusion
\cite{shlezinger,klafter}. When $<t>$ is finite, there is normal
diffusion: the mean-square displacement is $Dt$, where $D=\sigma
^{2}/6<t >$ for the three-dimensional case ($d=3$). Superdiffusion
takes place when the variance $\sigma ^{2}$ is infinite. Note that
in many of the superdiffusion realization cases, the decoupling
assumption of Eq. (\ref{dec}) can be inappropriate \cite{SWK}. In
what follows we consider only two regimes: normal diffusion and
subdiffusion.

\subsection{Integro-differential equations}

The interesting feature of the balance equations (\ref{n1}) and
(\ref{n2}) with $\Phi (s,\mathbf{z)}=\psi (s)\rho (\mathbf{z})$ is
that they can be rewritten as a system of integro-differential
equations:
\begin{equation}
\frac{\partial n_{1}}{\partial t}=\int_{0}^{t}\alpha
(t-s)\int_{R^{d}}\left[
n_{1}(s,\mathbf{x}-\mathbf{z})-n_{1}(s,\mathbf{x})\right] \rho
(\mathbf{z})d\mathbf{z}ds-\beta _{1}n_{1}+\beta _{2}n_{2}\, ,
\label{dn1}
\end{equation}
\begin{equation}
\frac{\partial n_{2}}{\partial
t}=f(n_{1},n_{2})+\beta_{1}n_{1}-\beta _{2}n_{2}\, ,  \label{dn2}
\end{equation}
where the memory kernel $\alpha (t)$ has to be determined. Let us
derive these equations from (\ref{n1}) and (\ref{n2}) by using the
Laplace transform for $\psi (t)$, and  the Fourier transform for
$\rho (\mathbf{x})$
\begin{equation}
\tilde{\psi}(H)=\mathcal{L}[\psi (t)]=\int_{0}^{\infty }
\psi(t)e^{-Ht}dt\,,\;\;\hat{\rho}(\mathbf{k})=\mathcal{F}[\rho
(\mathbf{x})]=\int_{R^{d}}\rho (\mathbf{x})e^{i\mathbf{k\cdot
x}}d\mathbf{x}\; \label{Lap}
\end{equation}
and the Fourier-Laplace (F-L) transform for the densities
$n_{k}(t,\mathbf{x})$
\begin{equation}
\hat{\tilde{n}}_{k}(H,\mathbf{k})=\mathcal{FL}[n_{k}(t,\mathbf{x}%
)]=\int_{R^{d}}\int_{0}^{\infty}n_{k}(t,\mathbf{x})e^{-Ht+i\mathbf{k\cdot
x} }dtd\mathbf{x}\,,\;\;\ k=1,2\, .  \label{FL}
\end{equation}
Equation (\ref{n1}) with $\Phi (s,\mathbf{z)}=\psi (s)\rho
(\mathbf{z})$ in the F-L space reads
\begin{eqnarray}
\hat{\tilde{n}}_{1}(H,\mathbf{k})
&=&\hat{n}_{1}(0,\mathbf{k})\frac{1-\tilde{\psi}(H+\beta_{1})}{H+\beta
_{1}}+\hat{\tilde{n}}_{1}(H,\mathbf{k})\hat{\rho}(\mathbf{k})
\tilde{\psi}(H+\beta _{1})  \notag \\
&&+\beta_{2}\hat{\tilde{n}}_{2}(H,\mathbf{k})\frac{1-
\tilde{\psi}(H+\beta_{1})}{H+\beta_{1}}\, . \label{La}
\end{eqnarray}
To perform the F-L transform in Eq. (\ref{FL}) we use the standard
convolution property
\begin{equation*}
\hat{\tilde{n}}_{1}(H,\mathbf{k})\hat{\rho}(\mathbf{k})\tilde{\psi}(H)=\int
\int_{0}^{\infty }\left[ \int_{0}^{t}\int
n_{1}(t-s,\mathbf{x}-\mathbf{z} )\rho (\mathbf{z})\psi
(s)d\mathbf{z}ds\right] e^{-Ht+i\mathbf{kx}}dtd \mathbf{x}\, .
\end{equation*}
Rearranging Eq. (\ref{La}) and introducing the `memory' kernel
$\alpha (t)$ in term of its Laplace transform:
\begin{equation}
\tilde{\alpha}(H)=\frac{\left( H+\beta _{1}\right)
\tilde{\psi}(H+\beta _{1}) }{\left(
1-\tilde{\psi}(H+\beta_{1})\right) }\, ,  \label{LaplaceMemory}
\end{equation}
we obtain
\begin{equation}
H\hat{\tilde{n}}_{1}(H,\mathbf{k})-\hat{n}_{1}(0,\mathbf{k})=
\hat{\tilde{n}}_{1}(H,\mathbf{k})\tilde{\alpha}(H)(\hat{\rho}(\mathbf{k})-1)+
\beta_{2}\hat{\tilde{n}}_{2}(H,\mathbf{k})-
\beta_{1}\hat{\tilde{n}}_{1}(H,\mathbf{k})\, . \label{11f}
\end{equation}
Applying the F-L transform inversion to Eq. (\ref{11f}), we obtain the
integro-differential equation (\ref{dn1}). To find the F-L
transform of Eq. (\ref{n2}), we denote the nonlinear proliferation
term by $Z(t,\mathbf{x})=f\left(
n_{1}(t,\mathbf{x}),n_{2}(t,\mathbf{x})\right)$. Its F-L transform
is
\begin{equation}
\hat{\tilde{Z}}(H,\mathbf{k})=\mathcal{LF}\left[
Z(t,\mathbf{x})\right] \, . \label{FLZ}
\end{equation}
We have from Eq. (\ref{n2})
\begin{equation}
\hat{\tilde{n}}_{2}(H,\mathbf{k})=\hat{n}_{2}(0,\mathbf{k})
\frac{1}{H+\beta_{2}}+\hat{\tilde{Z}}(H,\mathbf{k})\frac{1}{H+
\beta_{2}}+\beta_{1}\hat{\tilde{n}}_{1}(H,\mathbf{k})\frac{1}{H+\beta_{2}}\,
, \label{22}
\end{equation}
where $\hat{\tilde{Z}}(H,\mathbf{k})/(H+\beta _{2})=\mathcal{LF}
\int_{0}^{t}Z(t-s,\mathbf{x})e^{-\beta _{2}s}ds.$ Rearranging Eq.
(\ref{22}) in the following form
\begin{equation}
H\hat{\tilde{n}}_{2}(H,\mathbf{k})-\hat{n}_{2}(0,\mathbf{k})=\hat{\tilde{Z}}
(H,\mathbf{k})+\beta_{1}\hat{\tilde{n}}_{1}(H,\mathbf{k})-\beta_{2}\hat{
\tilde{n}}_{2}(H,\mathbf{k})\, ,  \notag
\end{equation}
and applying the F-L inversion and using Eq. (\ref{FLZ}), we
obtain Eq. (\ref{dn2}).

\subsection{Probability density function for cell jumps}

Now we are in a position to discuss different approximations for
the probability density function for cell jumps $\rho
(\mathbf{z})$. Of course this function is not symmetrical in
general. The cells of the migrating phenotype are biased to
migrate away from the tumor spheroid core. The reasons for this
asymmetrical creeping are the non-uniform nutrient concentration
(chemotaxis), the gradient of cell adhesion sites (haptotaxis),
etc. Experimental observations suggest that cell jumps are
controlled by adhesion of tumor cells to extracellular matrix and
jump lengths are very small \cite{Giese1}. Therefore $\rho
(\mathbf{z})$ is a rapidly decaying function for large
$|\mathbf{z|}$. In other words, the density of tumor cells varies
on the scales that are much larger than the typical jump length.
Thus one can use the Taylor series in Eq. (\ref{n1}) with $\Phi
(s,\mathbf{z)}=\psi (s)\rho (\mathbf{z})$ expanding
$n_{1}(t-s,\mathbf{x}-\mathbf{z})$ in $\mathbf{z}$ and truncate
the series at the $2$nd moment. This truncation for rapidly
decaying function $\rho (\mathbf{z})$ is a well defined procedure,
since the higher moments become progressively smaller
\cite{Murray}. We have
\begin{equation}
\int_{R^{d}}n_{1}(t-s,\mathbf{x}-\mathbf{z})\rho
(\mathbf{z})d\mathbf{z}=n_{1}(t-s,\mathbf{x})-<z_{i}>\frac{\partial
n_{1}}{\partial
x_{i}}+\frac{1}{2}<z_{i}z_{j}>\frac{\partial^{2}n_{1}}{\partial
x_{i}\partial x_{j}}+...\, , \label{expansion}
\end{equation}
where the Einstein rule for summation over repeated indices $i$
and $j$ is implied, and angular brackets denote averaging with
respect to $\rho(\mathbf{z}):$
\begin{equation}
<z_{i}>=\int_{R^{d}}z_{i}\rho (\mathbf{z})d\mathbf{z}\, ,
~~~~<z_{i}z_{j}>=
\int_{R^{d}}z_{i}z_{j}\rho(\mathbf{z})d\mathbf{z}\, .
\end{equation}
Substitution of Eq. (\ref{expansion}) into Eq. (\ref{n1}) with the
decouple property $\Phi (s,\mathbf{z)}=\psi (s)\rho (\mathbf{z})$
yields
\begin{eqnarray}
n_{1}(t,\mathbf{x}) &=&n_{1}(0,\mathbf{x})\Psi
(t)e^{-\beta_{1}t}+\int_{0}^{t}n_{1}(t-s,\mathbf{x)}\psi
(s)e^{-\beta_{1}s}ds-<z_{i}>\int_{0}^{t}\frac{\partial
n_{1}}{\partial x_{i}}\psi(s)e^{-\beta_{1}s}ds  \notag \\
+\frac{1}{2}
&<&z_{i}z_{j}>\int_{0}^{t}\frac{\partial^{2}n_{1}}{\partial
x_{i}\partial x_{j}}\psi
(s)e^{-\beta_{1}s}ds+\beta_{2}\int_{0}^{t}n_{2}(t-s,\mathbf{x)}\Psi
(s)e^{-\beta_{1}s}ds\, . \label{ex}
\end{eqnarray}
Note that the third term on the right hand side of this equation
reflects a bias of random walk in the direction $<\mathbf{z>.}$ In
fact, this equation involves the first two moments for random
jumps: $<z_{i}>$ and $<z_{i}z_{j}>.$ It can be rewritten as the
integro-differential equation
\begin{equation}
\frac{\partial n_{1}}{\partial t}+<z_{i}>\int_{0}^{t}\alpha
(t-s)\frac{\partial n_{1}}{\partial
x_{i}}ds=\frac{1}{2}<z_{i}z_{j}>\int_{0}^{t}\alpha
(t-s)\frac{\partial ^{2}n_{1}}{\partial x_{i}\partial
x_{j}}ds-\beta_{1}n_{1}+\beta _{2}n_{2}\, .  \label{mainmain}
\end{equation}
If the cell jumps are normally distributed then the characteristic
function of $\rho (\mathbf{z})$ is
\begin{equation}
\hat{\rho}(\mathbf{k})=\exp \left(
ia_{i}k_{i}-\frac{1}{2}\sigma_{ij}k_{i}k_{j}\right) \, ,
\end{equation}
where the summation convection is implied for the repeated index.
The positive definite matrix $\sigma _{ij}$ can be written in
terms of the first two moments
\begin{equation}
\sigma _{ij}=<z_{i}z_{j}\mathbf{>-}<z_{i}><z_{j}\mathbf{>.}
\end{equation}
The probability density function $\rho (\mathbf{z})$ is
\begin{equation}
\rho (\mathbf{z})=\frac{1}{\left( 2\pi \right)
^{d/2}(\mathrm{det\,}\sigma)^{1/2}} \exp \left(
-\frac{1}{2}\left(
\sigma ^{-1}\right) _{ij}(z_{i}-<z_{i}>)(z_{j}-<z_{j}>\right) ,
\end{equation}
where $\left( \sigma ^{-1}\right) _{ij}$ is an inverse matrix. If
we assume that there is no bias, $<z_{j}\mathbf{>=}0,$ and
$<z_{i}z_{j}>=0$ for $i\neq j$, and
$<z_{i}^{2}>=\frac{<z^{2}>}{d}=\frac{\sigma ^{2}}{d}$. Then Eq.
(\ref{ex}) takes the form
\begin{eqnarray}
n_{1}(t,\mathbf{x}) &=&n_{1}(0,\mathbf{x})\Psi (t)
e^{-\beta_{1}t}+\int_{0}^{t}n_{1}(t-s,\mathbf{x})\psi (s)e^{-\beta_{1}s}ds  \notag \\
&&+\frac{\sigma ^{2}}{2d}\int_{0}^{t}\Delta
n_{1}(t-s,\mathbf{x})\psi
(s)e^{-\beta_{1}s}ds+\beta_{2}\int_{0}^{t}n_{2}(t-s,\mathbf{x})\Psi
(s)e^{-\beta_{1}s}ds\, .
\end{eqnarray}
From the last equation one obtains integro-differential equation
for $n_{1}$ in $d$ dimension
\begin{equation}
\frac{\partial n_{1}}{\partial
t}=\frac{\sigma^{2}}{2d}\int_{0}^{t}\alpha (t-s)\Delta
n_{1}(s,\mathbf{x})ds-\beta_{1}n_{1}+\beta_{2}n_{2}\,.
\end{equation}
Note that the one-dimensional case ( $d=1)$ was analyzed in
\cite{fi07}.

\subsection{Memory kernel and waiting time probability density function}

The formula
\begin{equation}
\tilde{\alpha}(H)=\frac{\left( H+\beta_{1}\right)
\tilde{\psi}(H+\beta_{1})}{1-\tilde{\psi}(H+\beta_{1})}
\label{Lmemory}
\end{equation}
gives us the relationship between the transport memory kernel
$\alpha (t)$ in (\ref{mainmain}) and the waiting-time pdf $\psi
(t)$ in terms of their Laplace transforms. It should be emphasized
that it is impossible to find an explicit expression for the
memory kernel $\alpha (t)$ for arbitrary choices of the
waiting-time pdf $\psi (t)$. However, we are concerned with the
rate of the spreading of tumor cells. In what follows we show that
this rate depends on the Laplace transform $\tilde{\alpha}(H)$
rather than $\alpha (t)$. That is why the formula (\ref{Lmemory})
is so important for our analysis. It follows from (\ref{Lmemory})
that the transport kernel $\alpha (t)$ depends on the parameter
$\beta_{1}$. This means that we can not separate the transport
process and random switching in general. This phenomenon has been
discussed recently in the literature on anomalous transport with
reactions \cite{Helen,Hor}.

Let us consider three typical distributions for the waiting-time
pdf $\psi(t)$.

\textit{\ (i) Exponential distribution. }The random waiting time
is exponentially distributed if it has a density\textit{\ }
\begin{equation}
\psi (t)=\lambda e^{-\lambda t}\, .  \label{memo1}
\end{equation}
The Laplace transform for this distribution is
\begin{equation}
\tilde{\psi}(H)=\int_{0}^{\infty }\lambda e^{-\lambda
t}e^{-Ht}dt=\frac{\lambda }{\lambda +H}  \label{memo2}
\end{equation}
and
\begin{equation}
\tilde{\alpha}(H)=\frac{\left( H+\beta_{1}\right)
\tilde{\psi}(H+\beta_{1})}{\left(
1-\tilde{\psi}(H+\beta_{1})\right) }=\lambda \, ;  \label{memo3}
\end{equation}
therefore $\alpha (t)=\lambda \delta (t)$. In this case the kernel
$\alpha (t)$ is independent of $\beta_{1}.$ Thus we have a
classical system of convection-diffusion-reaction equations
\begin{equation}
\frac{\partial n_{1}}{\partial t}+v_{i}\frac{\partial
n_{1}}{\partial x_{i}}=D_{ij}\frac{\partial ^{2}n_{1}}{\partial
x_{i}\partial x_{j}}-\beta_{1}n_{1}+\beta_{2}n_{2}\, ,
\end{equation}
\begin{equation}
\frac{\partial n_{2}}{\partial
t}=f(n_{1,}n_{2})+\beta_{1}n_{1}-\beta_{2}n_{2}\, ,  \label{dif2}
\end{equation}
with the diffusion tensor $D_{ij}=\lambda <z_{i}z_{j}>/2$ and the velocity $%
\mathbf{v=}\lambda <\mathbf{z>}$.

\textit{\ (ii) Gamma distribution. } The waiting-time pdf $\psi
(t)$\ corresponds to the family of gamma distributions with
parameters $m$\ and $ \lambda $:
\begin{equation}
\psi (t)=\frac{\lambda ^{m}t^{m-1}e^{-\lambda t}}{\Gamma (m)}\,.
\end{equation}
Then $\tilde{\psi}(H)=\left( \frac{\lambda }{\lambda
+H}\right)^{m}$ and
\begin{equation}
\tilde{\alpha}(H)=\frac{\left( H+\beta_{1}\right)
\lambda^{m}}{(\lambda+H+\beta_{1})^{m}-\lambda ^{m}}\, .
\end{equation}
For example, if $m=2$
\begin{equation}
\tilde{\alpha}(H)=\frac{\lambda ^{2}\text{ }}{2\lambda
+H+\beta_{1}}\, ,
\end{equation}
and the memory kernel is
\begin{equation}
\alpha (t)=\lambda ^{2}e^{-(2\lambda +\beta_{1})t}\, .
\end{equation}
The main result here is that the transport memory kernel depends
on the parameter $\beta_{1}$. The integro-differential equation
for cells of migratory phenotype takes the form
\begin{equation}
\frac{\partial n_{1}}{\partial t}+v_{i}\lambda
\int_{0}^{t}e^{-(2\lambda +\beta_{1})s}\frac{\partial
n_{1}}{\partial x_{i}}ds=D_{ij}\lambda \int_{0}^{t}e^{-(2\lambda
+\beta_{1})s}\frac{\partial ^{2}n_{1}}{\partial x_{i}\partial
x_{j}}ds\mathbf{-}\beta_{1}n_{1}+\beta_{2}n_{2}\, .
\label{memory5}
\end{equation}
The integro-differential Eq. (\ref{memory5}) can be rewritten as
the hyperbolic reaction-transport equation, and corresponding
travelling wave solutions can be found as in \cite{FO,MFo} (see
also \cite{bbns2007}).

\textit{\ (iii) Power law waiting time distribution. } The power
law $\psi (t)\sim 1/(1+t/\tau )^{1+\gamma }$ with $0<\gamma <1$ is
used in many applications \cite{klafter}. Here we use $\tau$ which
is (in general case) not equal to $1/\lambda$ to stress the
fractional property of cell dynamics. It is more convenient to use
its Laplace transform
\begin{equation}
\tilde{\psi}(H)=\frac{1}{1+\left( H\tau \right) ^{\gamma }}.
\end{equation}
Then
\begin{equation}
\tilde{\alpha}(H)=\frac{\left( H+\beta_{1}\right) \tilde{\psi}(H+\beta_{1})%
}{\left( 1-\tilde{\psi}(H+\beta_{1})\right) }=\frac{\left(H
+\beta_{1}\right) ^{1-\gamma }}{\tau ^{\gamma }}\, .
\end{equation}

\section{ Cancer spreading rate}

The overall rate $u$ at which cancer cells spread is usually
defined as the velocity of the experimentally detectable tumor
front. In the generic Fisher equation setting the propagation rate
is $u=2\sqrt{DU}$, where $D$ is the diffusion coefficient and $U$
is the proliferation rate \cite{Murray}. The speed of this front
is determined by the processes taking place at the leading edge of
the cells' profile. In this paper we have a system of equations
(\ref{n1}) and (\ref{n2}) and we define the overall spreading rate
as the speed of the travelling wave solution of this system. For
front-like initial conditions, the fronts for both densities
$n_{1}$ and $n_{2}$ quickly achieve the stationary forms that
propagate with a constant rate $u$. The main purpose here is to
find the dependence of this propagation rate on the statistical
characteristics of the random switching process, $\beta_{1}$ and
$\beta_{2}$, two moments for random jumps: $\langle z_{i}\rangle$
and $\langle z_{i}z_{j}\rangle $ and waiting time distribution
$\psi (t)$. We use the logistic growth for cell proliferation
\begin{equation}
f(n_{1},n_{2})=Un_{2}\left( 1-(n_1+n_{2})/K\right)\, ,
\label{logistic}
\end{equation}
where $U$ is the cell proliferation rate and $K$ is the carrying
capacity of the environment. We assume that the initial tumor
spheroid of radius $R$ has the following distribution of cells
\begin{equation}
n_{k}(0,\mathbf{x})=\text{\ }\left\{
\begin{array}{l}
A_{k} \\
0
\end{array}
\begin{array}{l}
if\quad \sum_{i=1}^{d}x_{i}^{2}\leq R^{2}, \\
otherwise,
\end{array}
\right.
\end{equation}
where positive constant $A_{1}$ and $A_{2}$ represent the stable
equilibrium points of the densities $n_{1}$ and $n_{2}$. They can
be found from two equations $A_{1}+A_{2}=K$ and
$\beta_{1}A_{1}=\beta _{2}A_{2}:$
\begin{equation}
A_{1}=\frac{\beta_{2}K}{\beta_{1}+\beta_{2}}\, ,\quad
A_{2}=\frac{\beta_{1}K }{\beta_{1}+\beta_{2}}.
\end{equation}
We assume that the characteristic length scale for the tumor front
is much smaller than the radius of the initial tumor spheroid. We
also assume that the bias acts in the radial direction such that
$\langle\mathbf{z}\rangle=\langle r\rangle \mathbf{e}_{r}$. These
assumptions allow us to consider the propagation of the effective
plane front in the radial direction, neglecting all curvature
effects. We expect that the long time development leads to the
propagation of travelling fronts of permanent forms: $n_{1}\left(
r-ut\right) $ and $n_{2}\left( r-ut\right), $ where the rate $u$
is common to both densities $n_{1}$ and $n_{2}$.

The balance equations for densities $n_{1}$ and $n_{2}$ are of the form
\begin{eqnarray}
n_{1}(t,r) &=&n_{1}(0,r)\Psi (t)e^{-\beta_{1}t}+
\int_{0}^{t}n_{1}(t-s,r)\psi (s)e^{-\beta_{1}s}ds-<r>\int_{0}^{t}%
\frac{\partial n_{1}}{\partial r}\psi (s)e^{-\beta_{1}s}ds  \notag \\
&&+\frac{\sigma ^{2}}{2d}\int_{0}^{t}\frac{\partial ^{2}n_{1}}{\partial r^{2}%
}\psi (s)e^{-\beta_{1}s}ds+\beta_{2}\int_{0}^{t}n_{2}(t-s,r)\Psi
(s)e^{-\beta_{1}s}ds\, ,  \label{basic1}
\end{eqnarray}
\begin{eqnarray}
n_{2}(t,r)
&=&n_{2}(0,r)e^{-\beta_{2}t}+U\int_{0}^{t}n_{2}(t-s,r)(1-\left(
n_{1}(t-s,r)+n_{2}(t-s,r)\right) /K)e^{-\beta_{2}s}ds  \notag \\
&&+\beta_{1}\int_{0}^{t}n_{1}(t-s,r)e^{-\beta_{2}s}ds.
\label{basic2}
\end{eqnarray}
This system of equations is a starting point for the analysis of
plane front propagation in a radial direction.

\subsection{\protect\bigskip Hyperbolic scaling and Hamilton-Jacobi equation}

The objective here is to find the rate $u$ without resolving the
shape of the travelling waves \cite{F1,Fr}. For this purpose we
use a hyperbolic scaling $r\rightarrow r/\varepsilon
,\,t\rightarrow t/\varepsilon $ and the rescaled density
$n_{k}^{\varepsilon }\left( t,r\right) =n_{k}\left( t/\varepsilon
,r/\varepsilon \right) $ (see Appendix A). We write the density
$n_{k}^{\varepsilon }\left( t,r\right) $ in the exponential form
\begin{equation}
n_{k}^{\varepsilon }\left( t,r\right) =A_{k}\exp \left( -\frac{%
G^{\varepsilon }\left( t,r\right) }{\varepsilon }\right) ,\quad
k=1,2 \label{wkb}
\end{equation}
where the non-negative function $G^{\varepsilon }\left( t,r\right)
$ describes the logarithmic asymptotic of both densities and plays
a very important role. It follows from (\ref{wkb}) that as long as
the function
\begin{equation}
G\left( t,r\right) =\lim_{\varepsilon \rightarrow 0}G^{\varepsilon
}\left( t,r\right)
\end{equation}
is positive, the rescaled density $n_{k}^{\varepsilon }\left(
t,r\right) \rightarrow 0$ as $\varepsilon \rightarrow 0.$ We may
argue that the equation $G\left( t,r\left( t\right) \right) =0$
gives us the spreading front position $r\left( t\right) $ in the
long-time and large-distance limit \cite{F1}. Substitution of the
exponential transformation (\ref{wkb}) into the equations for the
rescaled densities $n_{i}^{\varepsilon }(t,r)$ and taking the
limit $\varepsilon \rightarrow 0$ yield two equations for $A_{1}$
and $A_{2}.$ These equations have a non-trivial solution when the
corresponding determinant is equal to zero (see Appendix A). It
gives the
following equation for $G\left( t,r\right) :$%
\begin{eqnarray}
&&\left[ 1-\left( 1+<r>\frac{\partial G}{\partial r}+\frac{\sigma
^{2}}{2d} \left( \frac{\partial G}{\partial r}\right) ^{2}\right)
\int_{0}^{\infty }e^{\frac{\partial G}{\partial t}s}\psi
(s)e^{-\beta_{1}s}ds\right] \left[ 1-U\int_{0}^{\infty
}e^{\frac{\partial G}{\partial t}s}e^{-\beta_{2}s}ds
\right]   \notag \\
&&-\beta_{1}\beta_{2}\int_{0}^{\infty }e^{\frac{\partial
G}{\partial t} s}\Psi (s)e^{-\beta_{1}s}ds\times \int_{0}^{\infty
}e^{\frac{\partial G}{\partial t}s}e^{-\beta_{2}s}ds=0.
\label{final}
\end{eqnarray}
In terms of the Laplace transform
$\tilde{\psi}(H)=\mathcal{L}[\psi (t) ]$, Eq. (\ref{final}) can be
rewritten as a generalized Hamilton-Jacobi equation
\begin{equation}
1-\left( 1+<r>\frac{\partial G}{\partial r}+\frac{\sigma
^{2}}{2d}\left( \frac{\partial G}{\partial r}\right) ^{2}\right)
\tilde{\psi}(-\frac{
\partial G}{\partial t}+\beta_{1})=\frac{\beta_{1}\beta_{2}(1-\tilde{\psi}
(-\frac{\partial G}{\partial t}+\beta_{1}))}{(-\frac{\partial
G}{\partial t} +\beta_{1})\left( -\frac{\partial G}{\partial
t}+\beta_{2}-U\right) }\, . \label{HJ}
\end{equation}
Note that inferring Eq. (\ref{HJ}), we do not make any assumptions
regarding waiting time pdf $\psi (s)$.

\subsection{Wavefront velocity}

Let us introduce the Hamiltonian function $H$ and the generalized
momentum $ p $
\begin{equation}
H=-\frac{\partial G}{\partial t}\, ,\text{ \ \ }p=\frac{\partial
G}{\partial r}\, . \label{H}
\end{equation}
Then Hamilton-Jacobi equation (\ref{HJ}) takes the form of the
quadratic equation:
\begin{equation}
<r>p+\frac{\sigma
^{2}p^{2}}{2d}-\frac{1}{\tilde{\psi}(H+\beta_{1})}\left[
1-\frac{\beta_{1}\beta_{2}(1-\tilde{\psi}(H+\beta_{1}))}{(H+\beta_{1})\left(
H+\beta_{2}-U\right) }\right] +1=0.  \label{mom}
\end{equation}
This equation is very important because it allows us to find the
spreading rate $u$ %%%%%%by using \cite{F1}
\begin{equation}
u=\frac{\partial H}{\partial p}=\frac{H}{p(H)}\, .
\label{HE}
\end{equation}
We may equivalently write $u=\min_{H}\left\{ \frac{H}{p(H)}\right\}$, so
$u=\frac{H}{p(H)}$, where $H$ can be found from equation
\begin{equation}
\frac{\partial p}{\partial H}=\frac{p(H)}{H}\label{eq51}
\end{equation}

Let us illustrate this formula by using the classical Fisher equation
\[\frac{\partial n}{\partial t}= D\frac{\partial^2 n}{\partial x^2}+Un(1-n)\]
for which the Hamiltonian is $H=Dp^2/2+U$. Using this expression, we obtain
\begin{equation}\label{eq52}
p(H)=\left(\frac{2H-2U}{D}\right)^{1/2}\, .
\end{equation}
From Eqs. (\ref{eq51}) and (\ref{eq52}) we obtain $H=Dp^2(H)=2U$, and therefore,
the spreading rate for the Fisher equation is $u_F=H/p(H)=2(DU)^{\frac{1}{2}}$.
This is the classical propagation speed.

In what follows we consider a case when the mean jump length in
the radial direction is zero, $<r>=0.$ If the random waiting time
is exponentially distributed (\ref{memo1}): $\psi (t)=\lambda
e^{-\lambda t}$, then the equation for the migratory cells is
\begin{equation}
\frac{\partial n_{1}}{\partial t}=D\frac{\partial
^{2}n_{1}}{\partial r^{2}} -\beta_{1}n_{1}+\beta_{2}n_{2}\,.
\label{dif1a}
\end{equation}
The momentum $p(H)$ can be found from (\ref{mom})
\begin{equation}
p^{2}=\frac{(H+\beta_{1})}{D}-\frac{\beta_{1}\beta_{2}}{D\left(
H+\beta_{2}-U\right) }.  \label{momen}
\end{equation}
If we assume that $\beta_{1}=\beta_{2}$, we can find from
(\ref{HE}) that $H=U$ and (\ref{momen}) $p=\left( U/D\right)
^{1/2}$, and $H=U$. Therefore, the spreading rate is $u_{0}=\left(
UD\right) ^{1/2}$ which is half of the classical Fisher-KPP
(Fisher-Kolmogorov-Petrovskii-Piskunov) propagation speed $u_F$.
This result shows that the propagation rate is independent of the
random migration-proliferation switching for
$\beta_{1}=\beta_{2}$. When $\beta_{1}\neq \beta_{2}$ one can find
the ratio of the propagation rate $u$ and $u_{0}=\left( UD\right)
^{1/2}$ as
\begin{equation}\label{nor}
\left( \frac{u}{u_{0}}\right) ^{2}=\frac{4(H+\beta_{2}-U)^3\left[
(H+\beta_{2}-U)(H+\beta_{1})-\beta_{1}\beta_{2}\right]}
{\left[(H+\beta_{2}-U)^2+\beta_{1}\beta_{2}\right]^2}\, ,
\end{equation}
where $H$ is determined by Eq. (\ref{eq51}).
For the fixed values of $\beta_1$ and $U$, the
wavefront propagation rate versus $\beta_2/\beta_1$ is depicted in Fig. 1.
\begin{figure}
\begin{center}
\epsfxsize=8cm
\leavevmode
    \epsffile{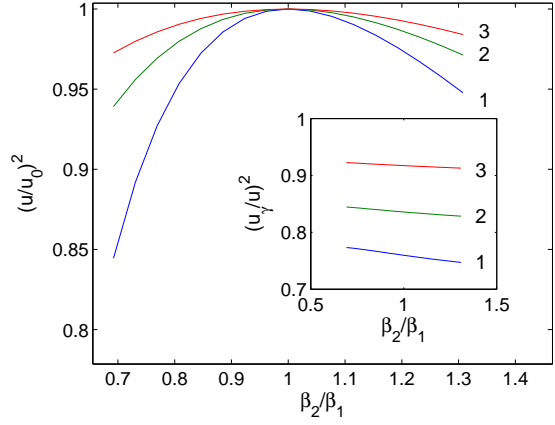}
\caption{(Color online) Propagation speed $\left(u/u_0\right)^2$
 vs  $\beta_2/\beta_1$. The values
of $\beta_1/U=[3, ~4.5, ~6.5]$ correspond to plots (1), (2), and
(3), respectively. The insert corresponds to a solution of Eqs.
(\ref{fr}) and (\ref{last}) for the same values of $\beta_1/U$ and
$\gamma=0.7$. }
\end{center}
\end{figure}

For the power law distribution $\psi (t)\sim (\tau /t)^{1+\gamma}$
with $0<\gamma <1$, the mean waiting time is divergent:
$<t>=\infty $. This assumption alone leads to the temporal
fractional differential operator and corresponding anomalous
diffusion equation \cite{klafter}. The mean squared displacement
for mobile cells is
\begin{equation}
<r^{2}(t)>=\frac{4D_{\gamma }}{\Gamma (1+\gamma )}t^{\gamma }\,
,\text{ } \label{an}
\end{equation}
where $D_{\gamma }=\sigma ^{2}/2d\tau ^{\gamma }.$

Let us find the overall propagation of cancer cells as a result of
interaction of subdiffusion (\ref{an}), logistic proliferation
(\ref{logistic}), and random migration-proliferation switching
(\ref{swit}). For the Laplace transform $\tilde{\psi}(H)=\left(
1+\left( H\tau \right) ^{\gamma }\right) ^{-1},$ the momentum
$p(H)$ can be found from (\ref{mom}):
\begin{equation}
p^{2}=\frac{(H+\beta _{1})^{\gamma }}{D_{\gamma
}}-\frac{\beta_{1}\beta_{2}(H+\beta _{1})^{\gamma -1}}{D_{\gamma
}\left( H+\beta_{2}-U\right) }\, . \label{fr}
\end{equation}
This formula together with (\ref{HE}) allows us to find the
overall propagation rate of tumor cells $u_{\gamma }$ for the
subdiffusion case. The case $\gamma =1$ corresponds to normal
diffusion. One can find from (\ref{HE}), (\ref{momen}) and
(\ref{fr}) the ratio of the anomalous propagation rate
$u_{\gamma}$ and the normal rate $u$ determined by (\ref{nor}):
\begin{equation}
\frac{u_{\gamma }}{u}=(H_{\gamma }\tau +\beta_{1}\tau
)^{\frac{1-\gamma }{2} }\, .  \label{last}
\end{equation}
Since the ``microscopic'' time $\tau $ is much smaller than the
characteristic ``cell proliferation'' time $U^{-1}$ and switching
time $\beta _{1}^{-1}$ and $H_{\gamma }\sim U$, we conclude that
$H\tau +\beta_{1}\tau <1$.  This condition of $H\sim U$ is also
confirmed by numerical solutions of Eqs. (\ref{eq51}),
(\ref{momen}), (\ref{fr}), and (\ref{last}) (see insert in Fig.
1). It follows from (\ref{last}) that the ratio $u_{\gamma }/u$
increases up to $1$ with $\gamma $ in the interval $0<\gamma <1$.
This means that normal diffusion leads to overestimation of the
overall cancer spreading. Note that the advantage of balance Eqs.
(\ref{n1}) and (\ref{n2}) is that they are related to a
``mesoscopic'' description of migratory cancer cells, and give us
the statistical meaning of the phenomenological reaction-transport
equation (\ref{mainmain}).

\section{Reaction-transport equations}

The influence of the migration-proliferation dichotomy on the
overall propagation rate is an important factor in glioma
development. The Markovian switching between two phenotypes
described by Eq. (\ref{swit}) can be generalized for the case when
memory effects are taken into account. The system of
integro-differential equations (\ref{dn1}) and (\ref{dn2}) takes
the form
\begin{eqnarray}
\frac{\partial n_{1}}{\partial t} &=&\int_{0}^{t}\alpha
(t-s)\int_{R^{d}} \left[
n_{1}(s,\mathbf{x}-\mathbf{z})-n_{1}(s,\mathbf{x})\right] \rho(
\mathbf{z})d\mathbf{z}ds  \notag \\
&&+\int_{0}^{t}[\mu _{2}(t-s)n_{2}(s,\mathbf{x})-\mu
_{1}(t-s)n_{1}(s,\mathbf{x})]ds\ ,  \label{intdef}
\end{eqnarray}
\begin{equation}
\frac{\partial n_{2}}{\partial
t}=f(n_{1},n_{2})-\int_{0}^{t}[\mu_{2}(t-s)n_{2}(s,\mathbf{x})
-\mu_{1}(t-s)n_{1}(s,\mathbf{x})]ds, \label{dn22}
\end{equation}
where $\mu_{i}\left( t\right) $ is the memory kernel for
non-Markovian switching. Combining Eqs. (\ref{intdef}) and
(\ref{dn22}) one finds that a total density $n=n_{1}+n_{2}$ obeys
the equation
\begin{equation}
\frac{\partial n}{\partial t}=\int_{0}^{t}\alpha
(t-s)\int_{R^{d}}\left[
n_{1}(s,\mathbf{x}-\mathbf{z})-n_{1}(s,\mathbf{x})\right] \rho
(\mathbf{z})d\mathbf{z}ds+f(n_{1},n_{2}).  \label{rte2}
\end{equation}
This equation does not restrict any possible random transitions
between migration and proliferation phenotypes. Moreover, it can
be a starting point of the glioma modelling in the framework of
the differential equations. It can be rewritten in terms of the
total density alone, if we introduce the probabilities $p_{j}$
such that $n_{1}=p_{1}n$ and $n_{2}=p_{2}n.$ By using the logistic
growth for cell proliferation $f(n_{1},n_{2})\equiv
f(n_2)=Un_{2}\left( 1-n_{2}/K\right) $ and rescaling
$p_2n\rightarrow n$, we obtain
\begin{equation}
\frac{\partial n}{\partial t}=p_{1}\int_{0}^{t}\alpha
(t-s)\int_{R^{d}}\left[
n(s,\mathbf{x}-\mathbf{z})-n(s,\mathbf{x})\right] \rho
(\mathbf{z})d\mathbf{z }ds+Up_{2}n\left( 1-n/K\right) .
\end{equation}

Let us find these probabilities for Markovian switching
(\ref{swit}). In fact there are four characteristic times in our
model: proliferation time ($ U^{-1}$ for logistic growth), the
transport time $<t>=\int_{0}^{\infty }t\psi (t)dt$ (averaging
waiting time), and two switching times $\beta_{1}^{-1}$ and $\beta
_{2}^{-1}.$ If we assume that both switching times are small
compared to the growth time $U^{-1}$ and transport time $<t>,$ the
``fast'' switching process can be averaged. The ``fast'' local
dynamics of densities $n_{1}$ and $n_{2}$ governed by the
equations
\begin{equation}
\frac{\partial n_{1}}{\partial
t}=-\beta_{1}n_{1}+\beta_{2}n_{2}\,,\quad \frac{\partial
n_{2}}{\partial t}=\beta_{1}n_{1}-\beta_{2}n_{2}\, .
\end{equation}
The solution for any $\mathbf{x}$ is
\begin{equation}
n_{1}(t)=\frac{\beta_{2}}{\beta_{1}+\beta_{2}}+\left[ n_{1}\left(
0\right) -\frac{\beta_{2}}{\beta_{1}+\beta_{2}}\right]
e^{-(\beta_{1}+\beta_{2})t}\, ,
\end{equation}
\begin{equation}
n_{2}(t)=\frac{\beta_{1}}{\beta_{1}+\beta_{2}}+\left[ n_{2}\left(
0\right) -\frac{\beta_{1}}{\beta_{1}+\beta_{2}}\right]
e^{-(\beta_{1}+\beta_{2})t}.
\end{equation}
For the large intermediate time $T$ such that $\beta _{1}^{-1}\sim
\beta_{2}^{-1}$ $\ll T\ll U^{-1}$, we have a local equilibrium,
that is, $n_{1}=\frac{\beta_{2}}{\beta_{1}+\beta_{2}}$ and
$n_{2}=\frac{\beta_{1}}{\beta_{1}+\beta_{2}}.$ If we consider now
the transport and proliferation, it is clear that the total number
of cancer cells $n$ splits locally to
$\frac{\beta_{2}}{\beta_{1}+\beta_{2}}n$ of migrating phenotype
and $\frac{\beta_{1}}{\beta_{1}+\beta_{2}}n$ of proliferating
phenotype. So
\begin{equation}
n_{1}(t,\mathbf{x})=\frac{\beta_{2}}{\beta_{1}+\beta_{2}}n(t,\mathbf{x}
),\quad
n_{2}(t,\mathbf{x})=\frac{\beta_{1}}{\beta_{1}+\beta_{2}}n(t,\mathbf{x})\,
.
\end{equation}
This means that we have only one variable $n(t,\mathbf{x})$ for
which we can formulate a balance equation considering the
transport for $n_{1}(t,\mathbf{x}) $ and proliferation for
$n_{2}(t,\mathbf{x})$. The probabilities are
\begin{equation}
p_{1}=\frac{\beta_{2}}{\beta_{1}+\beta_{2}},\quad
p_{2}=\frac{\beta_{1}}{ \beta_{1}+\beta_{2}}.
\end{equation}
In this limiting case, the model can be formulated in terms of the
linear balance equation for the total number of cancer cells per
unit volume $n(t,\mathbf{x})$
\begin{equation}
n(t,\mathbf{x})=\frac{\beta_{2}}{\beta_{1}+\beta_{2}}\int_{0}^{t}
\int_{R^{d}}n(t-s,\mathbf{x}-\mathbf{z})\psi (s)\rho
(\mathbf{z})d\mathbf{z}ds+\frac{\beta_{1}}{\beta_{1}+\beta_{2}}U\int_{0}^{t}n(t-s,\mathbf{x})ds.
\end{equation}
This reaction-transport equation can be also used to study the
wavefront propagation in the framework of the Hamiltonian-Jacobi
approach.

\section{Conclusion}

We developed a probabilistic approach for \textit{a
migration-proliferation dichotomy} in the spreading of tumor cells
in the invasive zone. We derived the balance equations for
densities of cancer cells of two phenotypes. In the migratory
state the cells randomly move but there is no cell proliferation,
while in the proliferating state the cancer cells do not migrate
and only proliferation takes place. We took into account random
switching between cell proliferation and migration by using a
two-state Markov chain. The transport of tumor cells is formulated
in terms of the CTRW with an arbitrary waiting time distribution,
while proliferation is modeled by a non-linear function of both
densities. We found the overall rate of tumor cell invasion for
both normal diffusion and subdiffusion. The advantage of our
probabilistic approach is that it allows us to take into account
anomalous (subdiffusive) transport within the general scheme of
migration, proliferation, and phenotype switching. We showed the
equivalence of balance equations to a system of
integro-differential equations involving memory effects for the
transport of mobile cells. By using a hyperbolic scaling and
Hamilton-Jacobi formalism we derived formulae for the overall
spreading rate of cancer cells. We showed that the memory effects
(subdiffusion) leads to a decrease in propagation rate compared to
a standard diffusion approximation for transport.

\section*{Acknowledgment}

Authors thank Daniel Campos, Werner Horsthemke and Vicenc Mendez \
for interesting discussions. This research was carried out under
the EPSRC grant EP/D03115X/1 and the Israel Science Foundation.

\section*{Appendix A}

Rescaling of Eqs. (\ref{basic1})\ and (\ref{basic2}), we obtain
\begin{eqnarray}
n_{1}^{\varepsilon }(t,r) &=&n_{1}^{\varepsilon }(0,r)\Psi
(t)e^{-\beta_{1}t/\varepsilon }+\int_{0}^{t/\varepsilon
}n_{1}^{\varepsilon }(t-\varepsilon s,r)\psi (s)e^{-\beta_{1}s}ds  \notag \\
&&- \varepsilon <r>\int_{0}^{t/\varepsilon }\frac{\partial
n_{1}^{\varepsilon }}{\partial r}\psi
(s)e^{-\beta_{1}s}ds+\frac{\varepsilon ^{2}\sigma ^{2}}{2d}
\int_{0}^{t/\varepsilon }\frac{\partial ^{2}n_{1}^{\varepsilon
}}{\partial
r^{2}}(t-\varepsilon s,r)\psi (s)e^{-\beta _{1}s}ds  \notag \\
&&+\beta_{2}\int_{0}^{t/\varepsilon }n_{2}^{\varepsilon
}(t-\varepsilon s,r)\Psi (s)e^{-\beta _{1}s}ds,
\end{eqnarray}
\begin{eqnarray}
n_{2}^{\varepsilon }(t,r) &=&n_{2}^{\varepsilon
}(0,r)e^{-\beta_{2}t/\varepsilon }+U\int_{0}^{t/\varepsilon
}n_{2}^{\varepsilon }(t-\varepsilon s,r)(1-\left(
n_{1}^{\varepsilon }+n_{2}^{\varepsilon
}\right) /K)e^{-\beta_{2}s}ds  \notag \\
&&+\beta_{1}\int_{0}^{t/\varepsilon }n_{1}^{\varepsilon
}(t-\varepsilon s,r)e^{-\beta _{2}s}ds.
\end{eqnarray}
Substitution of the exponential transformation $n_{k}^{\varepsilon
}\left(t,r\right) =A_{k}\exp \left( -\frac{G^{\varepsilon }\left(
t,r\right) }{\varepsilon }\right) $ into these equations and
accounting initial conditions yields
\begin{eqnarray}
A_{1} &=&A_{1}\int_{0}^{t/\varepsilon }\exp \left[
\frac{G^{\varepsilon }(t,r)-G^{\varepsilon }(t-\varepsilon
s,r)}{\varepsilon }\right] \psi(s)e^{-\beta _{1}s}ds  \notag \\
-\varepsilon  &<&r>A_{1}\exp \left( \frac{G^{\varepsilon }\left(
t,r\right) }{\varepsilon }\right) \int_{0}^{t/\varepsilon
}\frac{\partial }{\partial r} \exp \left( -\frac{G^{\varepsilon
}\left( t-\varepsilon s,r\right) }{\varepsilon }\right) \psi (s)e^{-\beta _{1}s}ds  \notag \\
&&+\frac{\varepsilon ^{2}\sigma ^{2}A_{1}}{2d}\exp \left( \frac{
G^{\varepsilon }\left( t,r\right) }{\varepsilon }\right)
\int_{0}^{t/\varepsilon }\frac{\partial ^{2}}{\partial r^{2}}\exp
\left( - \frac{G^{\varepsilon }\left( t-\varepsilon s,r\right)
}{\varepsilon }\right)
\psi (s)e^{-\beta_{1}s}ds  \notag \\
&&+\beta_{2}A_{2}\int_{0}^{t/\varepsilon }\exp \left[
\frac{G^{\varepsilon }(t,r)-G^{\varepsilon }(t-\varepsilon
s,r)}{\varepsilon }\right] \Psi (s)e^{-\beta_{1}s}ds,
\end{eqnarray}
\begin{eqnarray}
A_{2} &=&UA_{2}\int_{0}^{t/\varepsilon }\exp \left[
\frac{G^{\varepsilon }(t,r)-G^{\varepsilon }(t-\varepsilon
s,r)}{\varepsilon }\right] \left[ 1- \frac{A_{1}+A_{2}}{K}\exp
\left( -\frac{G^{\varepsilon }}{\varepsilon }
\right) \right] e^{-\beta_{2}s}ds  \notag \\
&&+\beta_{1}A_{1}\int_{0}^{t/\varepsilon }\exp \left[
\frac{G^{\varepsilon }(t,r)-G^{\varepsilon }(t-\varepsilon
s,r)}{\varepsilon }\right] e^{-\beta_{2}s}ds\, .
\end{eqnarray}
Taking the limit $\varepsilon \rightarrow 0$ we have
\begin{eqnarray}
A_{1} &=&A_{1}\int_{0}^{\infty }e^{\frac{\partial G}{\partial
t}s}\psi (s)e^{-\beta _{1}s}ds+A_{1}<r>\frac{\partial G}{\partial
r}\int_{0}^{\infty
}e^{\frac{\partial G}{\partial t}s}\psi (s)e^{-\beta _{1}s}ds  \notag \\
&&+\frac{\sigma ^{2}A_{1}}{2d}\left( \frac{\partial G}{\partial
r}\right)^{2}\int_{0}^{\infty }e^{\frac{\partial G}{\partial
t}s}\psi (s)e^{-\beta_{1}s}ds+\beta_{2}A_{2}\int_{0}^{\infty
}e^{\frac{\partial G}{\partial t} s}\Psi (s)e^{-\beta _{1}s}ds\, ,
\label{a11}
\end{eqnarray}
\begin{equation}
A_{2}=UA_{2}\int_{0}^{\infty }e^{\frac{\partial G}{\partial
t}s}e^{-\beta_{2}s}ds+\beta_{1}A_{1}\int_{0}^{\infty
}e^{\frac{\partial G}{\partial t}s}e^{-\beta _{2}s}ds. \label{a12}
\end{equation}
Then Eqs.(\ref{a11})\ and (\ref{a12}) can be rewritten as a system
of linear equations for $A_{1}$ and $A_{2}$
\begin{eqnarray}
&&A_{1}\left[ 1-\left( 1+<r>\frac{\partial G}{\partial
r}+\frac{\sigma ^{2}}{ 2d}\left( \frac{\partial G}{\partial
r}\right) ^{2}\right) \int_{0}^{\infty }e^{\frac{\partial
G}{\partial t}s}\psi (s)e^{-\beta _{1}s}ds\right]   \notag
\\
&&-A_{2}\beta _{2}\int_{0}^{\infty }e^{\frac{\partial G}{\partial
t}s}\Psi(s)e^{-\beta _{1}s}ds=0\, ,
\end{eqnarray}
\begin{equation}
A_{1}\beta _{1}\int_{0}^{\infty }e^{\frac{\partial G}{\partial
t}s}e^{-\beta_{2}s}ds-A_{2}\left[ 1-U\int_{0}^{\infty
}e^{\frac{\partial G}{\partial t} s}e^{-\beta _{2}s}ds\right] =0\,
.
\end{equation}
This system has a non-trivial solution when the corresponding
determinant is equal to zero:
\begin{eqnarray}
&&\left[ 1-\left( 1+<r>\frac{\partial G}{\partial r}+\frac{\sigma
^{2}}{2d} \left( \frac{\partial G}{\partial r}\right) ^{2}\right)
\int_{0}^{\infty }e^{ \frac{\partial G}{\partial t}s}\psi
(s)e^{-\beta _{1}s}ds\right] \left[1-U\int_{0}^{\infty
}e^{\frac{\partial G}{\partial t}s}e^{-\beta_{2}s}ds
\right]   \notag \\
&&-\beta_{1}\beta_{2}\int_{0}^{\infty }e^{\frac{\partial
G}{\partial t}s}\Psi (s)e^{-\beta _{1}s}ds\times \int_{0}^{\infty
}e^{\frac{\partial G}{\partial t}s}e^{-\beta _{2}s}ds=0\, .
\end{eqnarray}

\end{document}